\documentclass{aa}  
\usepackage{graphicx}
\usepackage{natbib}

\bibpunct{(}{)}{;}{a}{}{,}   % necessary for bibtex in A&A as well
\usepackage{txfonts}
\begin{document}
   \title{The high-mass star-forming region IRAS\,18182$-$1433}

%   \subtitle{I. Overviewing the $\kappa$-mechanism}

   \author{H.~Beuther\inst{1},
          Q.~Zhang\inst{2},
%       \fnmsep\thanks{Just to show the usage of the elements in the author field}
          T.K.~Sridharan\inst{2},
          C.-F.~Lee\inst{2}
         \and
          L.A.~Zapata\inst{2,3}
          }

   \offprints{H.~Beuther}

   \institute{Max-Planck-Institute for Astronomy, K\"onigstuhl 17, 
              69117 Heidelberg, Germany\\
              \email{beuther@mpia-hd.mpg.de}
         \and
              Harvard-Smithsonian Center for Astrophysics, 60 Garden Street,
              Cambridge, MA 02138, USA\\
             \email{name@cfa.harvard.edu}
         \and
             Centro de Radioastronom\'\i a y Astrof\'\i sica, UNAM, 
             Apdo. Postal 3-72 (Xangari), 58089 Morelia, Michoac\'an, M\'exico
%             \thanks{The university of heaven temporarily does not
%                     accept e-mails}
             }
\authorrunning{Beuther et al.}
\titlerunning{SMA observations of the HMPO IRAS\,18182$-$1433}

   \date{}

% \abstract{}{}{}{}{} 
% 5 {} token are mandatory 
  \abstract
    {}
    {We present mm line and continuum observations at high spatial
      resolution characterizing the physical and chemical properties
      of the young massive star-forming region IRAS\,18182$-$1433.}
    {The region was observed with the Submillimeter Array in the 1.3\,mm band.
      The data are complemented with short-spacing information from
      single-dish CO(2--1) observations. SiO(1--0) data from the VLA are added
      to the analysis.}
    {Multiple massive outflows emanate from the mm continuum peak. The
      CO(2--1) data reveal a quadrupolar outflow system consisting of two
      outflows inclined by $\sim$90$^{\circ}$. One outflow exhibits a
      cone-like red-shifted morphology with a jet-like blue-shifted
      counterpart where a blue counter-cone can only be tentatively
      identified. The SiO(1--0) data suggest the presence of a third outflow.
      Analyzing the $^{12}$CO/$^{13}$CO line ratios indicates decreasing CO
      line opacities with increasing velocities.  Although we observe a
      multiple outflow system, the mm continuum peak remains single-peaked at
      the given spatial resolution ($\sim$13500\,AU).  The other seven
      detected molecular species~-- also high-density tracers like CH$_3$CN,
      CH$_3$OH, HCOOCH$_3$~-- are all $\sim$1-$2''$ offset from the mm
      continuum peak, but spatially associated with a strong molecular outflow
      peak and a cm emission feature indicative of a thermal jet. This spatial
      displacement between the molecular lines and the mm continuum emission
      could be either due to an unresolved sub-source at the position of the
      cm feature, or the outflow/jet itself alters the chemistry of the core
      enhancing the molecular abundances toward that region.  A temperature
      estimate based on the CH$_3$CN$(12_k-11_k)$ lines suggests temperatures
      of the order 150\,K. A velocity analysis of the high-density tracing
      molecules reveals that at the given spatial resolution none of them
      shows any coherent velocity structure which would be consistent with a
      rotating disk. We discuss this lack of rotation signatures and attribute
      it to intrinsic difficulties to observationally isolate massive
      accretion disks from the surrounding dense gas envelopes and the
      molecular outflows.}
   {}
   
   \keywords{ stars: formation -- ISM: jets and outflows -- ISM: molecules --
     stars: early-type -- stars: individual (IRAS\,18182$-$1433) -- stars:
     winds, outflows }

   \maketitle
%
%________________________________________________________________

\section{Introduction}
\label{intro}

The formation of massive stars is an important topic by itself, but it
is also essential for understanding cluster formation and the Galactic
evolution. Significant observational progress has been made over the
last decade by statistical studies of large samples of young massive
star-forming regions (e.g.,
\citealt{plume1992,plume1997,molinari1996,molinari1998,molinari2000,hunter2000,sridha,beuther2002a,beuther2002b,beuther2002c,mueller2002,shirley2003,walsh2003,faundez2004,hill2005,klein2005}).
However, to really understand the physical properties at the inner
center of the massive star-forming regions at very early evolutionary
stages, high-spatial-resolution studies at (sub)mm wavelengths are a
more adequate way of investigation. Observational studies of various
massive star-forming regions with different interferometers have
started disentangling the complex nature of high-mass star formation
(e.g.,
\citealt{cesaroni1997,zhang1998a,wyrowski1999,shepherd2000,beuther2005c,gibb2003,garay2003,su2004}).
Nevertheless, we are still at the beginning of such studies, and there
is urgent need to increase the number of investigated sources at high
spatial and spectral resolution to base the conclusions on a more
statistical foundation. To mention a few specific questions we are
interested to tackle: What is the nature of massive molecular
outflows, in an evolutionary sense as well as between sources of
different mass and luminosity? What is the core mass distribution at
the earliest evolutionary stages, can we infer information about the
formation of the Initial Mass Function? What is the nature of
potentially embedded accretion disks, and which way to best
investigate them?  Here, we present Submillimeter Array
(SMA\footnote{The Submillimeter Array is a joint project between the
  Smithsonian Astrophysical Observatory and the Academia Sinica
  Institute of Astronomy and Astrophysics, and is funded by the
  Smithsonian Institution and the Academia Sinica.})  1.3\,mm
continuum and line observations toward the young high-mass
star-forming region IRAS\,18182$-$1433 addressing some of these
questions.

The source IRAS\,18182$-$1433 is part of an intensely studied sample
of 69 potential High-Mass Protostellar Objects (HMPOs) at early
evolutionary stages prior to forming a significant UCH{\sc ii} region
\citep{sridha,beuther2002a,beuther2002b,beuther2002c}. At a near
kinematic distance of 4.5\,kpc, the luminosity and gas mass from the
region are $\sim 10^{4.3}$\,L$_{\odot}$ and $\sim 1500$\,M$_{\odot}$,
respectively \citep{sridha,beuther2002a}. The region harbors a bipolar
outflow with an outflow mass of $\sim 30$\,M$_{\odot}$ and an outflow
energy of $\sim 2.3\times 10^{46}$\,erg \citep{beuther2002b}.
Furthermore, IRAS\,18182$-$1433 shows OH, Class {\sc ii} CH$_3$OH, and
H$_2$O maser emission \citep{forster1999,walsh1998,beuther2002c} but
was not detected in cm continuum emission down to 1\,mJy, indicating
the early evolutionary stage of the source \citep{sridha}. However,
recent sensitive VLA observations at cm and 7\,mm wavelengths detected
two weak cm and one 7\,mm continuum source in this region
\citep{zapata2006a}. A spectral index analysis of the three
sub-sources shows that the 7\,mm emission likely traces the main
massive protostellar object in the field, whereas one cm feature is
consistent with a thermal jet $\sim$2$''$ offset from the 7\,mm peak.
The other cm feature $\sim$11$''$ south-east of the 7\,mm peak shows
signatures of non-thermal synchrotron emission (\citealt{zapata2006a},
see also Fig.~\ref{continuum}).  \citet{debuizer2005} observed this
region in the mid-infrared at $10.5\,\mu$m and $18.1\,\mu$m
wavelength. They detected a weak mid-infrared source close to the
7\,mm and the cm thermal jet feature. Furthermore, they detect
another far stronger and extended mid-infrared source $\sim$10$''$
toward the south-east which appears to be associated with the cm
continuum emission with negative spectral index \citep{zapata2006a}.

\section{Observations}
\label{obs}

\subsection{Submillimeter Array Observations}

\begin{figure}[htb]
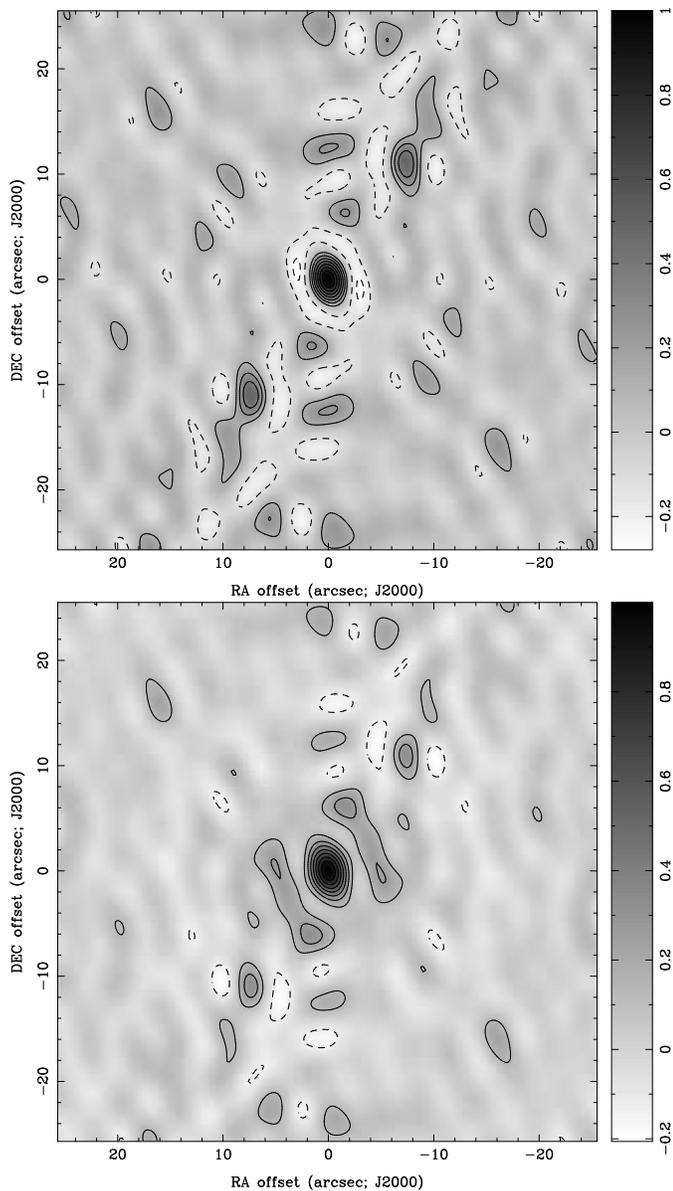

\includegraphics[angle=-90,width=8.8cm]{4887f1a.ps}\\
\includegraphics[angle=-90,width=8.8cm]{4887f1b.ps}
\caption{The dirty beams of the SMA data alone (top panel) and the
  merged SMA+30m data (bottom panel). The contour levels cover the range from
  $\pm 1$ in steps of $\pm 0.125$.}  \label{dirty} \end{figure}

The HMPO IRAS\,18182$-$1433 was observed with the Submillimeter Array
(SMA) on April 30th 2004 with 6 antennas at 1.3\,mm in the compact
configuration with projected baselines between 9.8 and
53\,k$\lambda$. The phase center was R.A.=18$^h$21$^m$09.098$^s$,
decl.=$-14^{\circ}31'49''.29$ (J2000.0) with a tuning frequency of
230.55\,GHz in the upper sideband ($v_{\rm{lsr}}=59.1$\,km\,s$^{-1}$). Parts
of the correlator were mis-functioning, narrowing down the usable
bandwidth to approximately 1.6\,GHz, the covered spectral range was
219.45 to 221.07\,GHz in the lower sideband (lsb) and 229.18 to
230.72\,GHz in the upper sideband (usb), respectively. The spectral
resolution was 0.8125\,kHz/channel corresponding to $\sim 1.1$\,km\,s$^{-1}$
velocity resolution. The weather was excellent with zenith opacities
$\tau(225\rm{GHz})$ between 0.1 and 0.07 measured by the National
Radio Astronomy Observatory (NRAO) tipping radiometer operated by the
Caltech Submillimeter Observatory (CSO). Passband and flux calibration
were derived from Uranus observations, and the flux density scale is
estimated to be accurate within 20\%. Phase and amplitude were
calibrated with regularly interleaved observations of the quasar
NRAO530 ($11.8^{\circ}$ from the source). We applied different
weightings for the continuum and line data resulting in synthesized
beams between $3.6''\times 2.4''$ and $3.9''\times 2.5''$. The rms
noise of the line and continuum images is dominated by the side-lobes
and the corresponding inadequate cleaning. The dirty beam with
side-lobes at approximately a 40\% level is shown in
Fig.~\ref{dirty}. The $3\sigma$ rms of the 1.3\,mm continuum image is
$\sim 15$\,mJy. The varying rms of the different line images is
discussed at the appropriate places in sections \ref{spectra} \&
\ref{outflow}. The initial flagging and calibration was done with the
IDL superset MIR originally developed for the Owens Valley Radio
Observatory \citep{scoville1993} and adapted for the SMA\footnote{The
MIR cookbook by Charlie Qi can be found at
http://cfa-www.harvard.edu/$\sim$cqi/mircook.html.}. The imaging and
data analysis were conducted in MIRIAD \citep{sault1995}.

\subsection{Short spacings from the IRAM 30\,m}
\label{shortspacings}

To complement the CO(2--1) observations with the missing short spacing
information, we combined the SMA data with the previously published
CO(2--1) map from the IRAM 30\,m telescope \citep{beuther2002b}. For
observing details of the single-dish observations we refer to the
paper by \citet{beuther2002b}. The single-dish data were
converted to visibilities within the MIRIAD package (using the task
UVMODEL). The single-dish and interferometer data were then
subsequently processed together. The weighting of the two datasets was
chosen to recover large-scale emission but maintain at the same time
the high-spatial resolution to resolve small-scale structure.
Convolving the merged dataset to the $11''$ beam of the 30\,m data
alone, the peak fluxes in the channel map are at approximately a 70\%
level of the single-dish measured fluxes. Measuring the integrated
fluxes toward the same outflow regions for the merged dataset and the
single-dish data alone, we even recover in the merged dataset
$\sim$88\% and $\sim$89\% of the blue and red CO(2--1) single-dish
emission, respectively. The resulting dirty beam from the merged
dataset is shown in Fig.~\ref{dirty} as well. The synthesized beam of
the combined data is $4.2''\times 2.8''$ (P.A.~15$^{\circ}$).

\section{Results}

\subsection{Millimeter continuum emission}
\label{cont}

\begin{figure}[htb] %\begin{center}
\includegraphics[angle=-90,width=8.8cm]{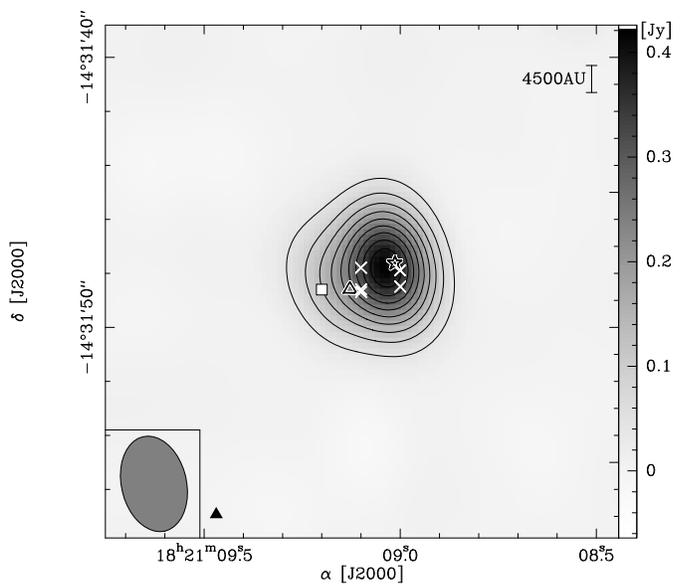} %\end{center}
\caption{1.3\,mm continuum image of IRAS\,18182$-$1433. The contours
range from 10 to 90\% (step 10\%) from the peak flux shown in Table
\ref{cont_para}. The synthesized beam ($3.6''\times
2..4''$\,P.A. 13.9$^{\circ}$) is shown at the bottom-left. The star
marks the 7\,mm source and the open and filled triangles the cm
sources with positive and negative spectral index, respectively
\citep{zapata2006a}. The square and white crosses mark the positions
from the CH$_3$OH Class {\sc ii} and H$_2$O masers, respectively
\citep{walsh1998,beuther2002c}.}  \label{continuum} \end{figure}

\begin{figure}[htb]
%\begin{center}
\includegraphics[angle=-90,width=8.8cm]{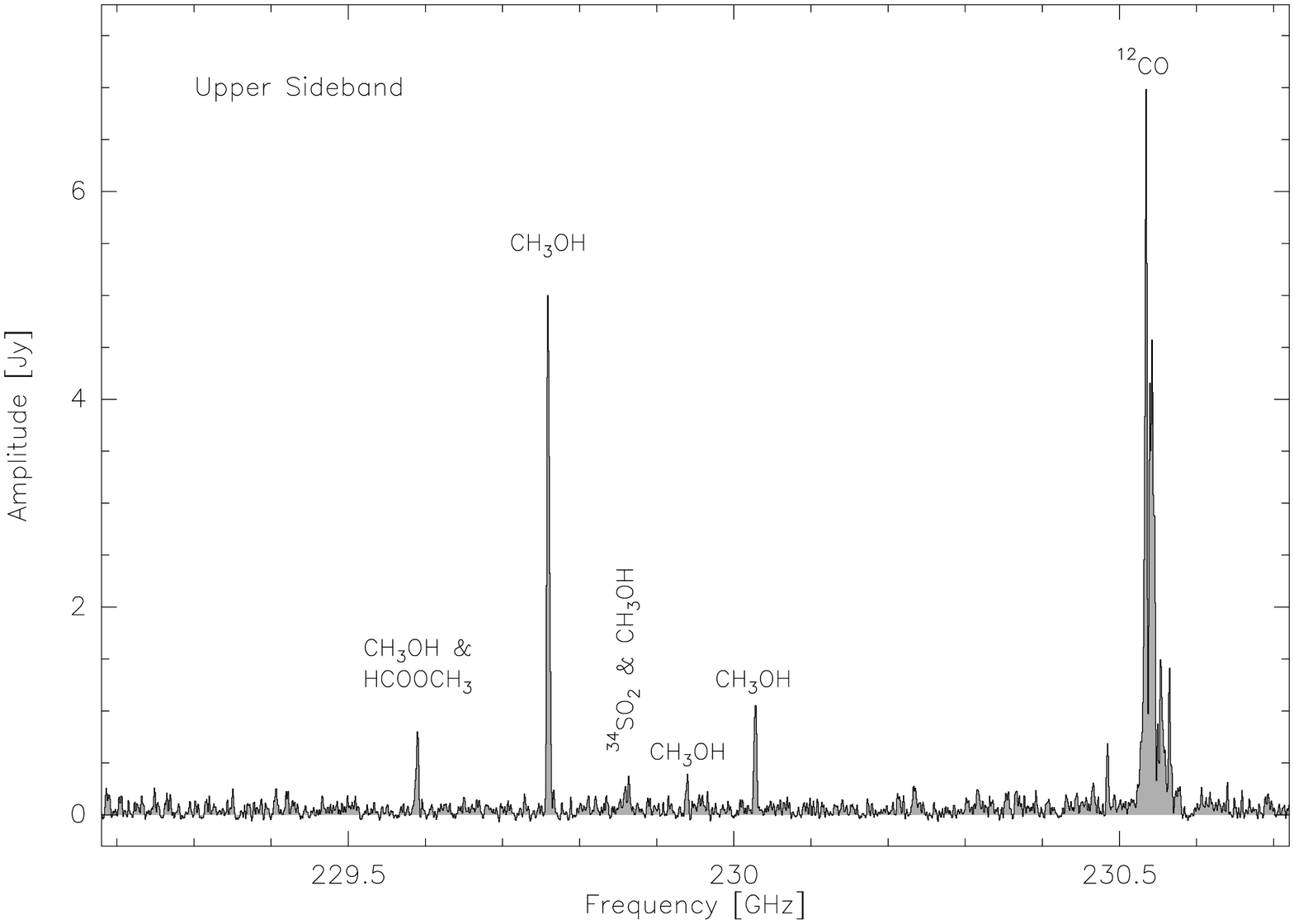}\\
\includegraphics[angle=-90,width=8.8cm]{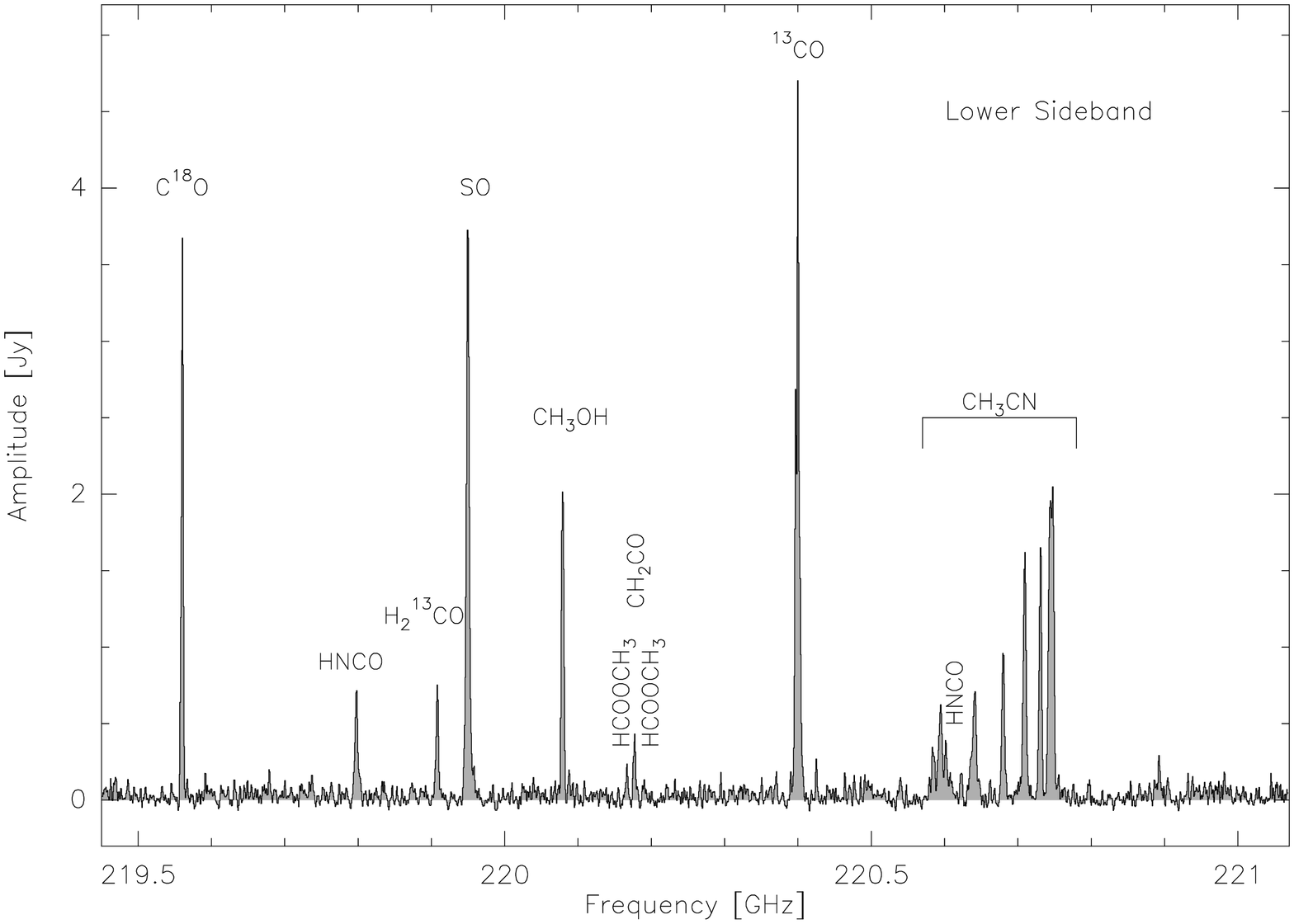}
%\end{center}
\caption{Upper and lower sideband spectra vector-averaged over all 
baselines. The molecular lines are marked.}
\label{uvspectra}
\end{figure}

Figure \ref{continuum} presents the synthesized 1.3\,mm continuum image of the
region produced by averaging the apparently line-free parts of the upper and
lower sideband spectra presented in Figure \ref{uvspectra}. We identify only
one massive isolated source at the given resolution and sensitivity. The 7\,mm
continuum source from \citet{zapata2006a} is associated with the 1.3\,mm peak
position, whereas the cm source with positive spectral index~-- thus likely a
thermal jet feature~-- is approximately $2''$ to the south-east
\citep{zapata2006a} and appears to be associated with the elongation of the
1.3\,mm emission in this direction. The CH$_3$OH Class {\sc ii} maser feature
\citep{walsh1998} is probably rather associated with the cm jet feature than
with the mm continuum peak. H$_2$O maser emission shows various maser features
distributed over the region \citep{beuther2002c}, some of them appear to be
associated with the cm feature whereas others are found closer to the mm
continuum peak.  The weak mid-infrared feature detected by
\citet{debuizer2005} may be more closely associated with the cm feature than
with the mm peak.  However, the given mid-infrared astrometry ($\sim$1.5$''$)
is not good enough to clearly associate that feature with one or the other
peak, and we refrain from further interpretation of the mid-infrared and mm/cm
continuum emission. Although the spatial resolution of the 1.3\,mm continuum
data is insufficient, the continuum extension in the direction of the cm jet
feature indicates that the mm continuum emission is not only of protostellar
nature but that there can be additional mm continuum emission due to an
underlying jet (for other similar examples see \citealt{gueth2003} or
\citealt{beuther2004e}).  Nevertheless, the majority of the mm continuum
emission is due to cold dust from the protostellar object(s).

\begin{table}
\caption{Millimeter Continuum Data \label{cont_para}}
\begin{center}
\begin{tabular}{lr}
\hline \hline
$S_{\rm{peak}}$ & 426\,mJy/beam \\ $S_{\rm{int}}$ & 546\,mJy \\
Mass@43K & 47.6\,M$_{\odot}$ \\ $N_{\rm{H_2}}$@43K &
5.7\,$10^{23}$cm$^{-2}$\\ Mass@150K & 12.4\,M$_{\odot}$ \\
$N_{\rm{H_2}}$@150K & 1.5\,$10^{23}$cm$^{-2}$\\
\hline \hline
\end{tabular}
\end{center}
\end{table}

Compared with the single-dish flux at 1.2\,mm wavelength of 3.9\,Jy
\citep{beuther2002a}, the integrated interferometer flux of $\sim
546$\,mJy is rather low, of the order 13\% from the single-dish
measurements. Since these interferometer observations are not
sensitive to spatial scales $>25''$, we filter out a rather smoothly
distributed large-scale halo, and only the compact core remains.
Assuming optically thin dust emission, we can estimate the mass and
column density of this central 1.3\,mm dust and gas core following the
equations outlined in \citet{hildebrand1983} and \citet{beuther2002a}.
We calculate these parameters assuming two temperatures, 43\,K as
derived from two-components grey-body fits \citep{sridha}, and 150\,K
consistent with the CH$_3$CN$(12_k-11_k)$ emission (see \S
\ref{ch3cn}). Based on previous interferometric observations of
similar sources from the original sample (IRAS\,18089$-$1732 \&
IRAS\,20293+3952, \citealt{beuther2004b,beuther2004e}), we use a dust
opacity index $\beta$ of 1, corresponding to a dust opacity per unit
dust mass of $\kappa (1.3\rm{mm}) \sim 1.8$\,cm$^2$\,g$^{-1}$. We note
that $\beta =1.4$ corresponds to $\kappa (1.3\rm{mm}) \sim
0.9$\,cm$^2$\,g$^{-1}$, as suggested by \citealt{ossenkopf1994}, and
using a larger $\kappa$ as we do results in lower mass estimates.
Depending on the temperature, the central core mass is between 12.4
and 47.6\,M$_{\odot}$. The H$_2$ column density varies between
$1.5\times 10^{23}$ and $5.7\times 10^{23}$\,cm$^{-2}$ (Table
\ref{cont_para}), corresponding to visual extinctions of a few hundred
($A_v=N_{\rm{H_2}}/0.94\times 10^{21}$, \citealt{frerking1982}). At a
temperature of 43\,K, the $3\sigma$ rms of 15\,mJy corresponds to a
mass sensitivity of $\sim 1.3$\,M$_{\odot}$.

We tried to derive radial intensity profiles, but since the core is
only marginally resolved and the data suffer strongly from the missing
short spacings, the resulting intensity distribution is unreasonably
steep prohibiting a further analysis in this direction.

\subsection{Spectral line emission}
\label{spectra}

Figure \ref{uvspectra} presents the observed spectral bandpass around
220 and 230\,GHz in the lsb and usb, respectively. Altogether we
detect 25 lines from 8 species ($^{12}$CO, HNCO, H$_2^{13}$CO, SO,
$^{34}$SO$_2$, CH$_3$OH, HCOOCH$_3$, CH$_3$CN) and 2 additional CO
isotopologues ($^{13}$CO and C$^{18}$O) with lower energy levels
$E_{\rm{lower}}$ between 5 and 568\,K (Table \ref{lines}). The broad
range of energy levels shows the power of broad spectral bandpasses
which allow to study very different excitation regimes
simultaneously.

\begin{figure*}[htb] \begin{center}
\includegraphics[angle=-90,width=17.6cm]{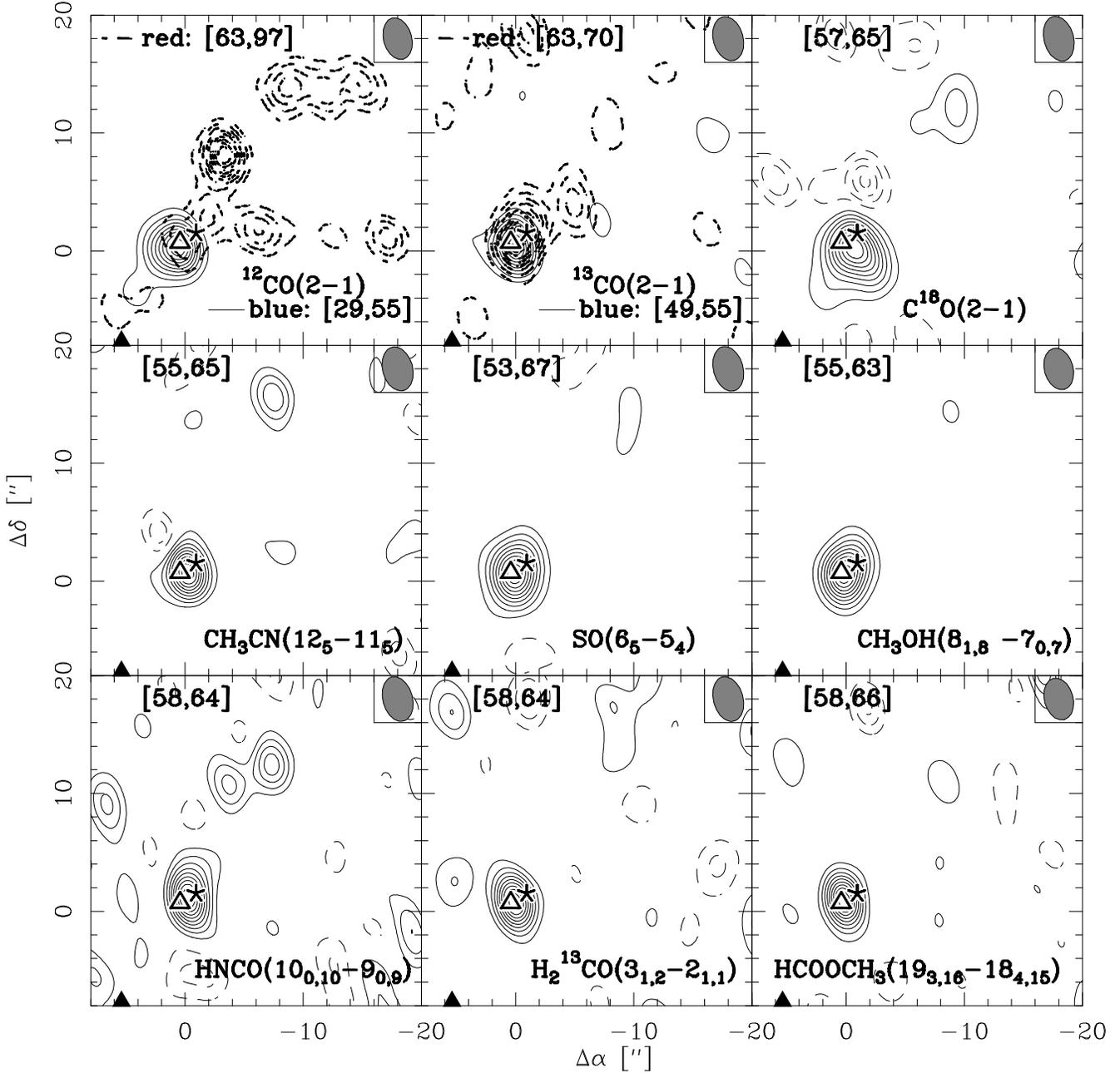} \end{center}
\caption{Integrated images of the observed molecular species. The full
  and dashed contours present the integrated emission as labeled in
  $^{12}$CO and $^{13}$CO image. The full and dashed contours in the
  remaining panels present the positive and negative (due to the
  side-lobe emission and inadequate cleaning) integrated emission as
  labeled in the panels. The contour levels range from 10 to 90\%
  (step 10\%) of the peak emission reported in Table \ref{peak}. There
  we also list the $1\sigma$ rms of each integrated image. The star
  marks the 1.3\,mm continuum peak position from Fig.~\ref{continuum},
  the open and full triangle mark the cm emission sources with
  positive and negative spectral index, respectively
  \citep{zapata2006a}. The synthesized beams are shown at the
  top-right of each panel.}
\label{images} \end{figure*}

Figure \ref{images} presents integrated line images of
all species and isotopologues (except $^{34}$SO$_2$ which is blended
with CH$_3$OH and too weak for imaging). For $^{12}$CO and $^{13}$CO
we are showing the blue and red line wing emission identifying a
bipolar outflow centered on the mm continuum peak. While the blue
$^{12}$CO emission is confined to a small region south-east of the
mm-peak, the red $^{12}$CO emission shows cone-like emission toward
the north-west. The outflow will be discussed in detail in \S
\ref{outflow}. All other spectral lines show compact emission close to the mm
emission peak. Interestingly none of these species peak exactly toward
the mm continuum, but all peak positions are shifted a little bit to
the south-east (of the order $1-2''$), largely coinciding with the cm
emission peak attributed to the thermal jet
\citep{zapata2006a}. Although the shift in the spectral line maps is
smaller than the angular resolution of the synthesized beam, we
consider it a real observational feature because (a) the line and
continuum images are extracted from the same dataset, and (b) by fitting
peak positions one can achieve a significant higher angular resolution
than the actual beam size HPBW depending on the signal to noise ration
S/N (down to $0.5\times \rm{HPBW/(S/N)}$, \citealt{reid1988}). A close
inspection of the higher density tracing $^{13}$CO blue and red
outflow emission shows that both line wings shows strong emission
toward the south-east of the mm peak as well, and are thus also
associated with the cm jet. A spatial offset between most
molecular species~-- particularly high-density tracers like CH$_3$CN
or HCOOCH$_3$~-- and the mm continuum emission is surprising, and
various potential origins will be discussed in \S \ref{discussion}.

\begin{table}
\caption{Observed lines}
\label{lines}
\begin{center}
\begin{tabular}{lrr}
\hline \hline
$\nu$ & line & $E_{\rm{lower}}$ \\
$[$GHz$]$ &      & $[$K$]$ \\
\hline
LSB\\
\hline
219.560 & C$^{18}$O$(2-1)$ & 5.3 \\
219.798 & HNCO$(10_{0,10}-9_{0,9})$ & 48 \\
219.909 & H$_2^{13}$CO$(3_{1,2}-2_{1,1})$ & 22 \\
219.949 & SO$(6_5-5_4)$ & 24 \\
220.079 & CH$_3$OH$(8_{0,8}-7_{1,6})$ & 85 \\
220.167 & HCOOCH$_3(17_{2,15}-16_{4,2})$ & 93 \\
220.178 & CH$_2$CO$(1_{1,11}-10_{1,10})$ & 66 \\
220.190 & HCOOCH$_3(17_{4,13}-16_{4,12})$ & 93 \\
220.399 & $^{13}$CO$(2-1)$ & 5.3 \\
220.585 & HNCO$(10_{1,9}-9_{1,8})$ & 91\\
220.594 & CH$_3$CN$(12_6-11_6)$ & 315 \\
220.641 & CH$_3$CN$(12_5-11_5)$ & 237 \\
220.679 & CH$_3$CN$(12_4-11_4)$ & 173 \\
220.709 & CH$_3$CN$(12_3-11_3)$ & 123 \\
220.730 & CH$_3$CN$(12_2-11_2)$ & 87 \\
220.743 & CH$_3$CN$(12_1-11_1)$ & 65 \\
220.747 & CH$_3$CN$(12_0-11_0)$ & 58 \\
\hline
USB\\
\hline
229.589 & CH$_3$OH$(15_{4,11}-16_{3,13})$E & 362 \\
229.590 & HCOOCH$_3(19_{3,16}-18_{4,15})$E & 106 \\
229.759 & CH$_3$OH$(8_{1,8}-7_{0,7})$E & 77 \\
229.858 & $^{34}$SO$_2(4_{2,2}-3_{1,3})$ & 7.7 \\
229.864 & CH$_3$OH$(19_{5,15}-20_{4,16})$A+ & 568 \\
229.939 & CH$_3$OH$(19_{5,14}-20_{4,17})$A$-$ & 568 \\
230.027 & CH$_3$OH$(3_{2,2}-4_{1,4})$E & 28 \\
230.538 & $^{12}$CO$(2-1)$ & 5.5 \\
\hline \hline
\end{tabular}
\end{center}
\end{table}

\begin{table}
\caption{Line image parameters to Fig.~\ref{images}}
\label{peak}
\begin{center}
\begin{tabular}{lrr}
\hline \hline
line & $1\sigma$ & $S_{\rm{peak}}$ \\
 & [$\frac{\rm{mJy}}{\rm{beam}}\frac{\rm{km}}{\rm{s}}$] & [$\frac{\rm{Jy}}{\rm{beam}}\frac{\rm{km}}{\rm{s}}$] \\
\hline
$^{12}$CO$(2-1)$ blue             & 1040 & 37.7 \\
$^{12}$CO$(2-1)$ red              & 1088 & 29.2 \\
$^{13}$CO$(2-1)$ blue             & 192  & 4.2  \\
$^{13}$CO$(2-1)$ red              & 224  & 5.1  \\
C$^{18}$O$(2-1)$                  & 440  & 13.3 \\
CH$_3$CN$(12_5-11_5)$             & 200  & 4.6  \\
SO$(6_5-5_4)$                     & 630  & 22.0 \\
CH$_3$OH$(8_{1,8}-7_{0,7})$E      & 480  & 24.9 \\
HNCO$(10_{0,10}-9_{0,9})$         & 240  & 2.4  \\
H$_2^{13}$CO$(3_{1,2}-2_{1,1})$   & 168  & 2.8  \\
HCOOCH$_3(19_{3,16}-18_{4,15})$E  & 224  & 4.3  \\
\hline \hline
\end{tabular}
\end{center}
\end{table}

\subsection{The molecular outflows}
\label{outflow}

\begin{figure}[htb]
\begin{center}
\includegraphics[angle=-90,width=8.8cm]{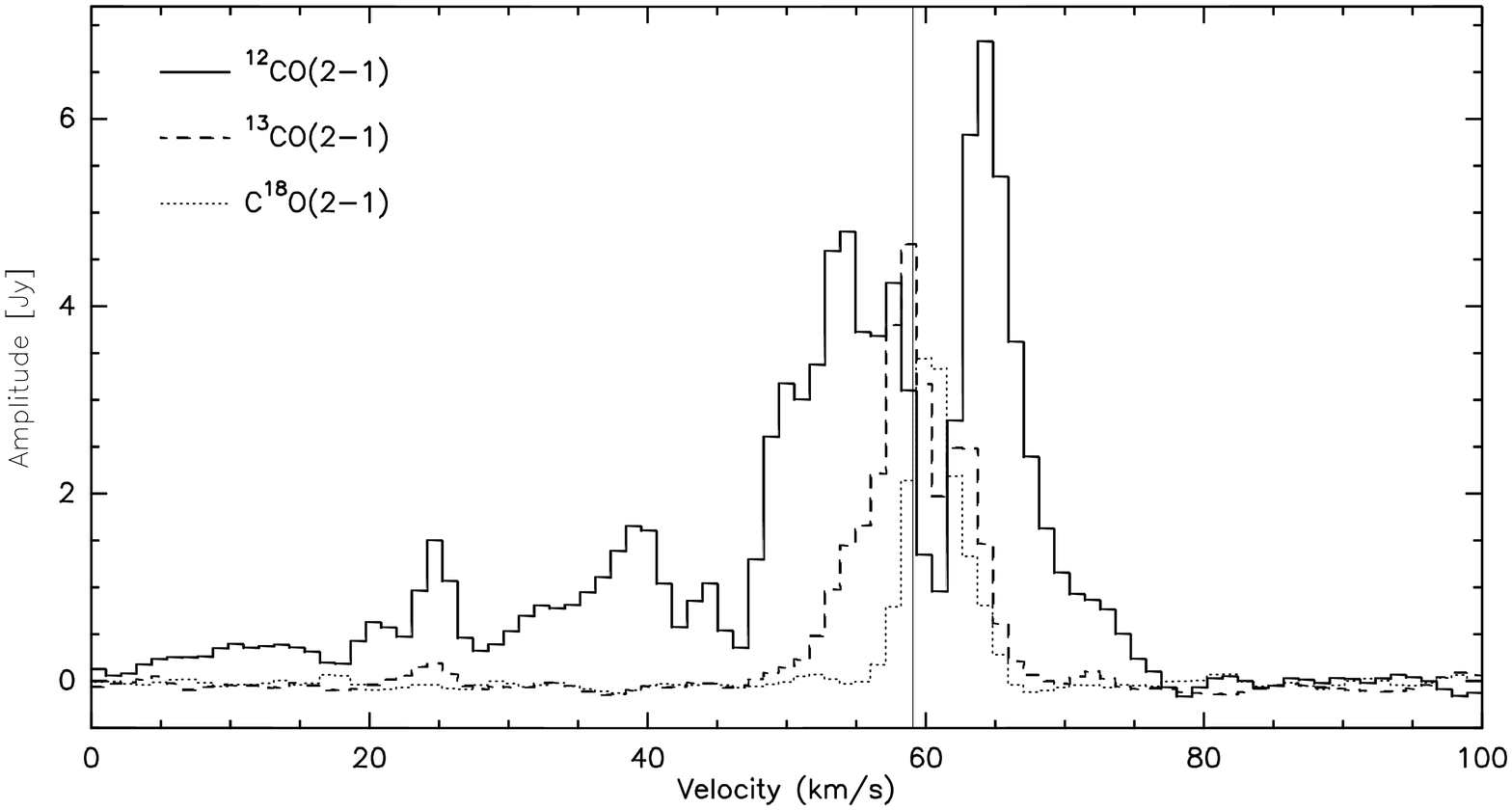}
\end{center}
\caption{SMA-only $^{12}$CO, $^{13}$CO and C$^{18}$O(2--1) spectra
  vector-averaged over all baselines. The spectral resolution is the
  nominal resolution of 1.1\,km\,s$^{-1}$. The $v_{\rm{lsr}}$ at
  59.1\,km\,s$^{-1}$ is marked by a vertical line.}
\label{uvspectra_co}
\end{figure}

Figure \ref{uvspectra_co} shows the SMA-only $^{12}$CO, $^{13}$CO and
C$^{18}$O spectra vector-averaged over all baselines. The
$^{12}$CO spectrum shows a dip around the systemic velocity
$v_{\rm{lsr}}$ of $59.1$\,km\,s$^{-1}$, which is not real absorption
but caused by the missing short-spacing information in the
interferometer data. The interferometric spatial filtering affects the
most abundant isotopologue $^{12}$CO strongest because it is optically
thick at the $v_{\rm{lsr}}$, tracing the extended cloud envelope. In
contrast to that, the less abundant isotopologues $^{13}$CO and
C$^{18}$O have lower opacities, they are tracing more compact
structures and hence do not show the pronounced dip at the
$v_{\rm{lsr}}$.  Nevertheless, already these vector-averaged spectra
in the uv-domain show $^{12}$CO high-velocity line wings due to the
molecular outflow between 30 and 100\,km\,s$^{-1}$.  Although weaker
and spanning a narrower velocity range, the line-wing emission is also
found in the $^{13}$CO data but barely discernable in the rare
isotopologue C$^{18}$O.

\begin{figure*}[htb] \begin{center}
\includegraphics[angle=-90,width=17.6cm]{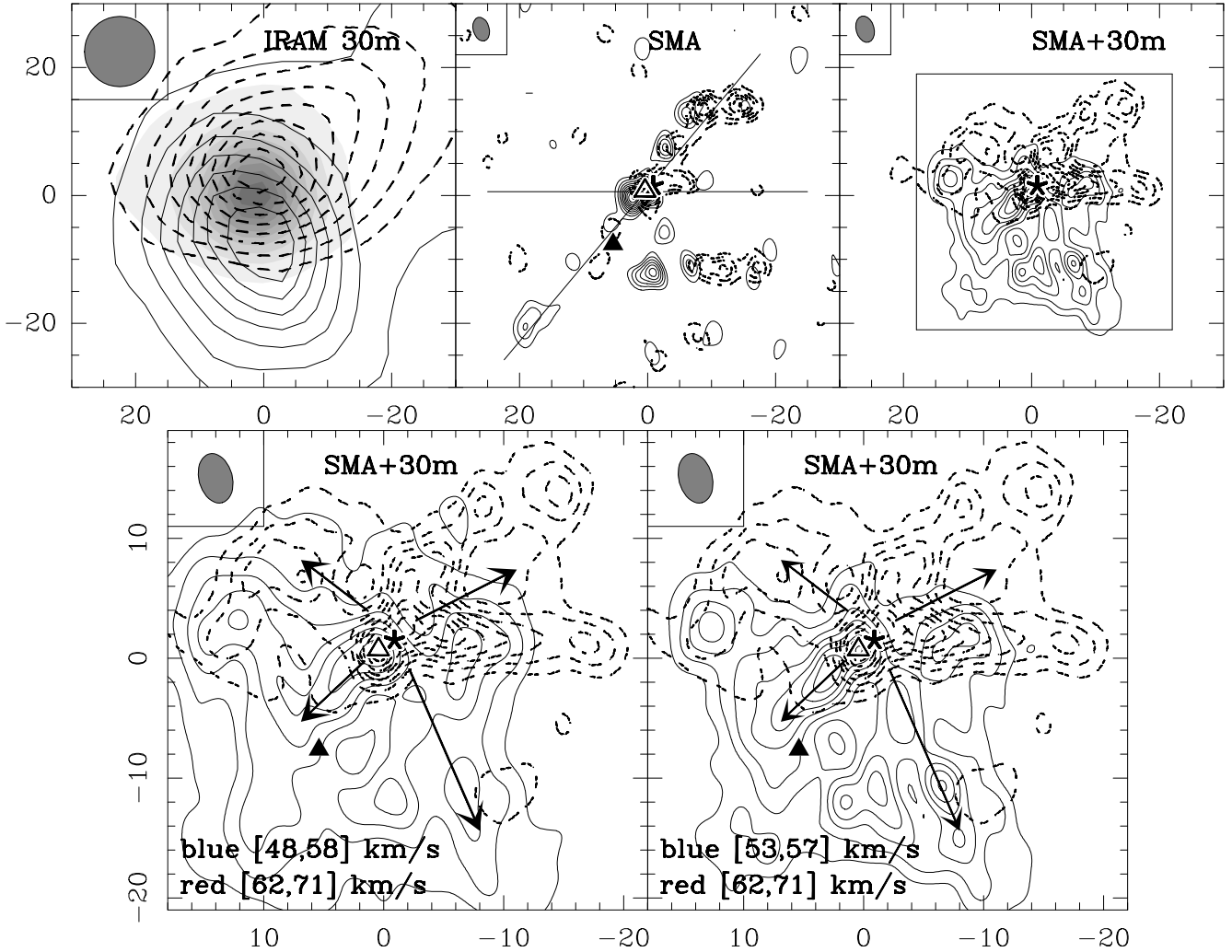} \end{center}
\caption{$^{12}$CO(2--1) outflow images observed with the IRAM 30\,m
  telescope \citep{beuther2002b} and the SMA. The top-left panel shows
  the 30\,m data only, the top-middle presents the SMA data only. The
  top-right and the bottom panels then show the combined SMA+30\,m
  data. The velocity-integration regimes for the top-row panels and
  the bottom-right panel are: blue (full contours)
  [53,57]\,km\,s$^{-1}$ and red (dashed contours)
  [62,71]\,km\,s$^{-1}$. The bottom-left panel has a slightly broader
  blue velocity range from [48,58]\,km\,s$^{-1}$. The grey-scale in
  the top-left panel presents the 1.2\,mm continuum emission from the
  IRAM 30\,m \citep{beuther2002a}. The stars in the other panels mark
  the mm continuum peak from the SMA observations. The open and full
  triangles in the top-middle and bottom panels mark the positions of
  the cm emission peaks with positive and negative spectral index,
  respectively \citep{zapata2006a}. The (synthesized) beams are shown
  at the top-left of each panel, and the axes are offsets in
  arcseconds from the phase center. The lines in the top-middle panel
  show the pv-cuts presented in Fig.\ref{pos_velo}, and the arrows in
  the bottom panel mark the approximate directions of the outflows
  discussed in the main text. The contour levels are listed in Table
  \ref{contours}.}
\label{sma_and_30m} \end{figure*}

Since we have the corresponding $^{12}$CO(2--1) single-dish
observations from the IRAM 30\,m telescope (\citealt{beuther2002b} and
Fig.~\ref{sma_and_30m} top-left panel), we were able to combine
interferometer and single-dish data and thus recover the missing short
spacings. The IRAM 30\,m data were less sensitive than the new SMA
observations detecting high-velocity emission only approximately
between 50 and 70\,km\,s$^{-1}$. Therefore, we confined the SMA data
to the range between 47.5 and 72.5\,km\,s$^{-1}$ to combine both
datasets.  Figure \ref{channel_co} presents a $^{12}$CO(2--1) channel
map showing the merged SMA+30\,m data between 50 and 70\,km\,s$^{-1}$
and the SMA only data at higher velocities relative to the
$v_{\rm{lsr}}$. Figure \ref{sma_and_30m} shows the resulting outflow
images covering all spatial scales from the single-dish to the
interferometer observations.

\begin{figure*}[htb] \begin{center}
\end{center} \caption{$^{12}$CO(2--1) channel map with 5\,km\,s$^{-1}$
  resolution. The channels between 35 to 45 km\,s$^{-1}$ and between
  75 to 90\,km\,s$^{-1}$ are produced from SMA only data, whereas the
  channels inbetween are from the merged SMA+30M dataset (labeled in
  each panel with the corresponding central velocity). For the SMA
  only data, we plot the contour levels from $\pm
  0.5$\,Jy\,beam$^{-1}$km\,s$^{-1}$ in $\pm
  0.5$\,Jy\,beam$^{-1}$km\,s$^{-1}$ steps. The $1\sigma$ rms of the
  weak emission channel at 90\,km\,s$^{-1}$ is
  0.15\,Jy\,beam$^{-1}$km\,s$^{-1}$. The SMA+30m data are contoured
  differently in each panel to account for the dynamic range
  differences. The start- and step-values are the same in units of
  Jy\,beam$^{-1}$km\,s$^{-1}$; 50\,km\,s$^{-1}$: $\pm 1.6$;
  55\,km\,s$^{-1}$: $\pm 4.4$; 60\,km\,s$^{-1}$: $\pm 6.6$;
  65\,km\,s$^{-1}$: $\pm 3.4$; 70\,km\,s$^{-1}$: $\pm 1.0$. The
  crosses mark the position of the mm continuum peak, and the open and
  full triangles mark the positions of the cm sources with positive
  and negative spectral index \citep{zapata2006a}.}
\label{channel_co}
\end{figure*}

\begin{figure}[tb]
\begin{center}
\includegraphics[angle=-90,width=8.8cm]{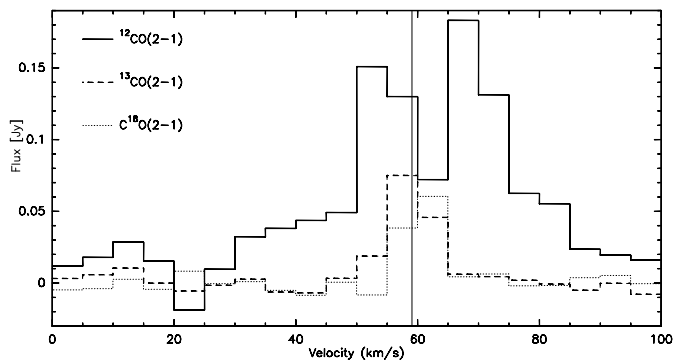}
\end{center}    
\caption{SMA-only $^{12}$CO, $^{13}$CO and C$^{18}$O(2--1) spectra
  extracted from the data cube after imaging with a spectral
  resolution of 5\,km\,s$^{-1}$. The spectra are averaged spatially
  over the whole region covered by the CO outflow (see top-left panel
  in Fig.~\ref{images}). The $v_{\rm{lsr}}$ at 59.1\,km\,s$^{-1}$ is
  marked by a vertical line.}
\label{spectra_co_images}
\end{figure}

The interferometer data alone exhibit CO emission approximately
between 30 and 100\,km\,s$^{-1}$ (Figs.~\ref{channel_co} \&
\ref{spectra_co_images}). We tried to image the 24\,km\,s$^{-1}$
feature (Fig.~\ref{uvspectra_co}) but could not identify any clear
signature associated with IRAS\,18182$-$1433.  Therefore, we conclude
that this spectral feature has to be associated with an unrelated
fore- or background cloud. The red-shifted feature in the SMA-only CO
outflow image in Figure \ref{sma_and_30m} (top-middle panel) at
Declination $\sim -11''$ and R.A. between approximately $-5''$ and
$-20''$ is an artifact of the image deconvolution and the strong
side-lobes caused by the insufficient uv-coverage and missing
short-spacings of the SMA data alone. After adding in the short
spacing information from the IRAM 30\,m observations, this feature is
largely gone, whereas the southern edge of the red-shifted CO cone at
Declination $0''$ is recovered much better (Figs.~\ref{channel_co} \&
\ref{sma_and_30m}).

We find red and blue outflow emission between 30 and 70\,km\,s$^{-1}$
south-east of the mm peak, whereas toward the north-west only red CO emission
is detected.  The north-western emission exhibits a clear cone-like
morphology, whereas the south-eastern emission appears jet-like with the
compact component being associated with the thermal jet feature observed in cm
emission \citep{zapata2006a}. Integrating over slightly different velocity
regimes for the blue wing, the observed structures do vary a bit: Employing
the originally from the single-dish data alone identified blue range from 53
to 57 km\,s$^{-1}$, the south-eastern jet-feature dominates the blue wing of
this outflow. However, integrating a little bit broader regime from 48 to 58
km\,s$^{-1}$, one can tentatively identify associated structures at offset
($10''/-13''$) and ($12''/3''$) which could be part of a blue counter-cone.
This potential blue cone-like morphology is less prominent than for the red
north-western wing, and with this dataset we cannot unambiguously decide
whether the blue outflow wing only shows jet-like emission or whether it
actually features a cone-like morphology.

\begin{table}
\caption{Line image parameters to Fig.~\ref{sma_and_30m}}
\label{contours}
\begin{center}
\begin{tabular}{lr}
\hline \hline
& Contours \\
 & [Jy\,beam$^{-1}$km\,s$^{-1}$] \\
\hline
30m blue & 66(22)198\\
30m red & 41(21)185 \\
SMA blue & 1.36(1.36)12.24\\
SMA red & 2.07(2.07)18.63\\
SMA+30m blue & 10.92(3.60)32.76 \\
SMA+30m red & 8.82(4.41)39.69 \\
\hline \hline
\end{tabular}
\end{center}
\end{table}

The single-dish data alone indicated an outflow oriented roughly in
north-south direction whereas the interferometer data suggest an
orientation approximately in north-west south-east direction. The
combined data now show that we are likely dealing with a system of
multiple outflows: The one outflow (a) is the north-west south-east
outflow revealed in detail by the SMA observations. However, most of
the single-dish flux at blue-shifted velocities is to the south of the
mm continuum peak, and Figure \ref{sma_and_30m} shows additional
red-shifted emission toward the north-east of it. One possibility
would be that the whole blue/red outflow structure is due to one
inhomogeneous wide-angle outflow with its symmetry axis along the jet
direction. The features observed with the SMA would then potentially
trace the denser parts of the molecular outflow. In contrast to this
picture, the cone-like red-shifted emission as well as the jet-like
blue-shifted features are strongly suggestive of at least one more
homogeneous outflow in north-west south-east direction as discussed
above. To account for the additional blue-shifted emission in the
south and the red-shifted emission toward the north-east, we interpret
these features as due to a second outflow (b) emanating from the same
mm continuum core.  The two outflows are inclined approximately
90$^{\circ}$ to each other (projected on the plane of the sky). The
corresponding blue and red outflow lobes are for both outflows not
exactly in opposite directions (as sketched in
Fig.~\ref{sma_and_30m}), but appear to be bend a little bit. Since we
expect that the two outflows are interacting with each other, such an
observed bending is no surprise but can be attributed to overlapping
contributions from all outflows. In summary, the merged CO(2--1)
dataset indicate a quadrupolar outflow system emanating from the main
mm continuum source in IRAS\,18182$-$1433.

\begin{figure}[h] \begin{center}
\includegraphics[angle=-90,width=8.8cm]{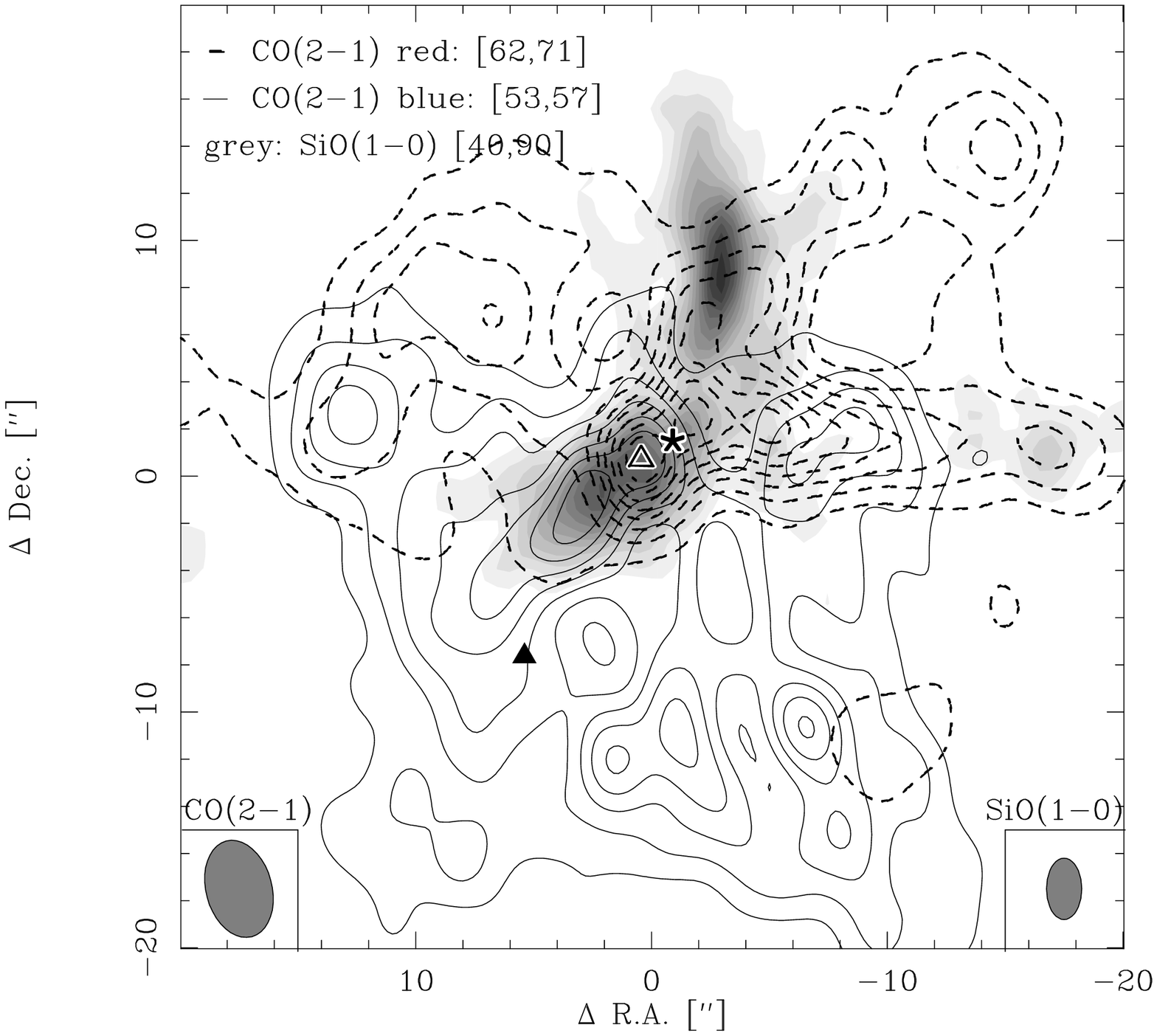} \end{center}
\caption{The thin full and dashed contours show the same blue and red
  CO(2--1) observations merged from the SMA and IRAM\,30\,m data as
  presented in Fig.~\ref{sma_and_30m}. The grey-scale shows
  the VLA SiO(1--0) emission integrated from 40 to 90\,km\,s$^{-1}$
  (kindly provided by Yuan Chen et al., in prep.). The contour levels
  of the CO images are presented in Table \ref{contours}, and the SiO
  data are contoured from 18 to 342\,Jy\,beam$^{-1}$km\,s$^{-1}$ (step
  27\,Jy\,beam$^{-1}$km\,s$^{-1}$). The CO and SiO synthesized beams
  are shown at the bottom-left and bottom-right, respectively. The
  cross marks the position of the mm continuum peak, and the open and
  full triangles mark the positions of the cm sources with positive
  and negative spectral index, respectively \citep{zapata2006a}.}
\label{co_sio}
\end{figure}

To increase the complexity of the outflow system from
IRAS\,18182-1433, Fig.~\ref{co_sio} presents an overlay of the
CO(2--1) data from the SMA+30m observations with VLA SiO(1--0)
observations integrated over the velocity range from 40 to
90\,km\,s$^{-1}$ (kindly provided by Yuan Chen et al., in prep.). The
south-eastern SiO feature is consistent with the blue jet-like CO
emission, whereas the SiO emission north of the mm continuum peak
shows a different morphology compared to the red CO cone-like outflow.
The strong northern SiO peak position appears to be correlated with
the northern edge of the CO cone in that direction. However, the rest
of the SiO emission continues even further north, unrelated to the CO
outflow. We detect additional weaker SiO emission at an offset
($-16''/1''$) spatially associated with the western end of the CO
cone. The additional SiO peak at ($-16''/1''$) and the northern peak
correlated with the northern CO cavity wall allow the interpretation
that the SiO emission may be due to the same outflow observed in CO,
just more strongly excited in the cavity walls.  However, the SiO
extension further north is suggestive of an additional third outflow
so far not detected in the CO emission.

\begin{figure}[htb] \begin{center}
\includegraphics[angle=-90,width=7.5cm]{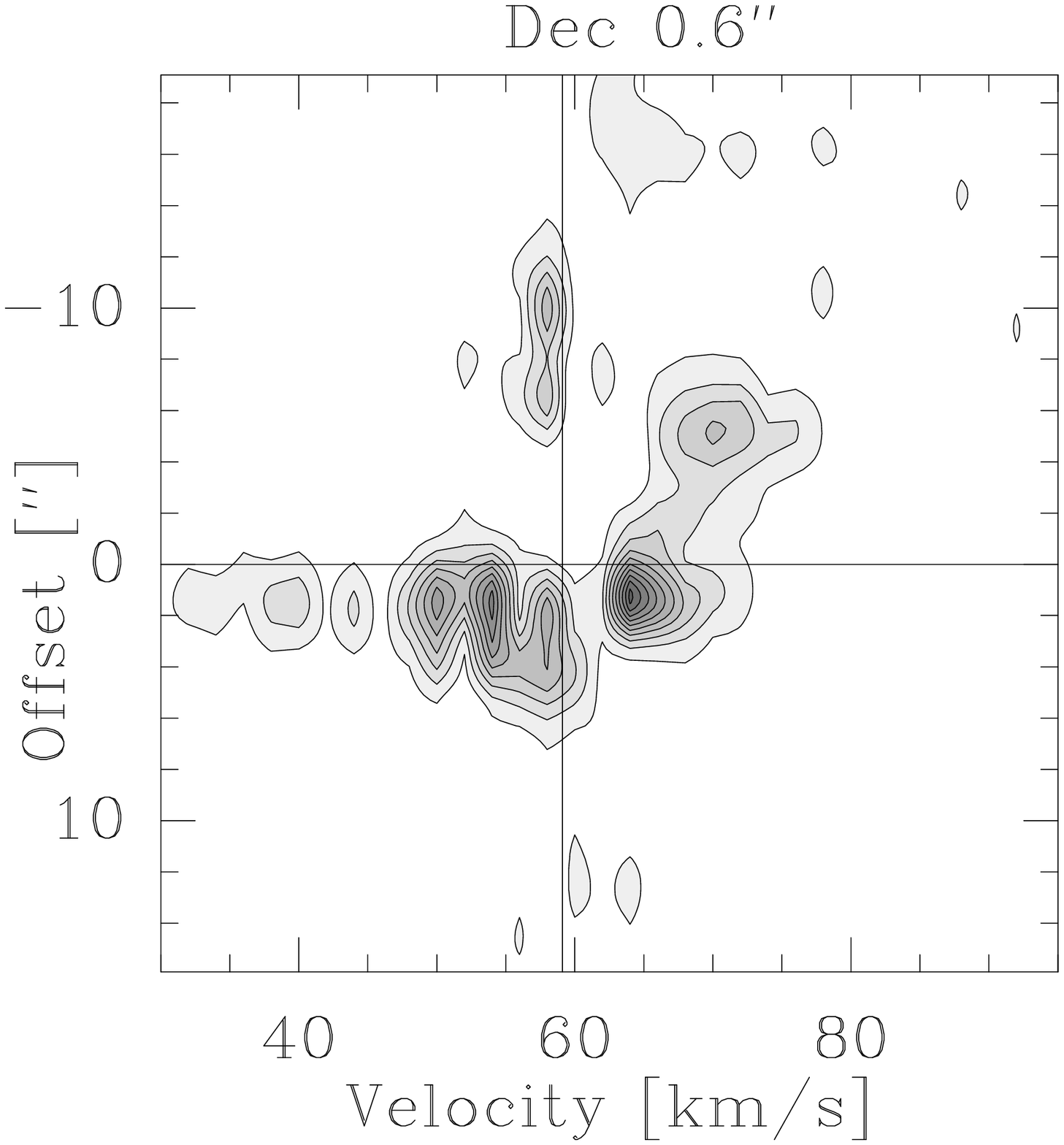}
\includegraphics[angle=-90,width=7.5cm]{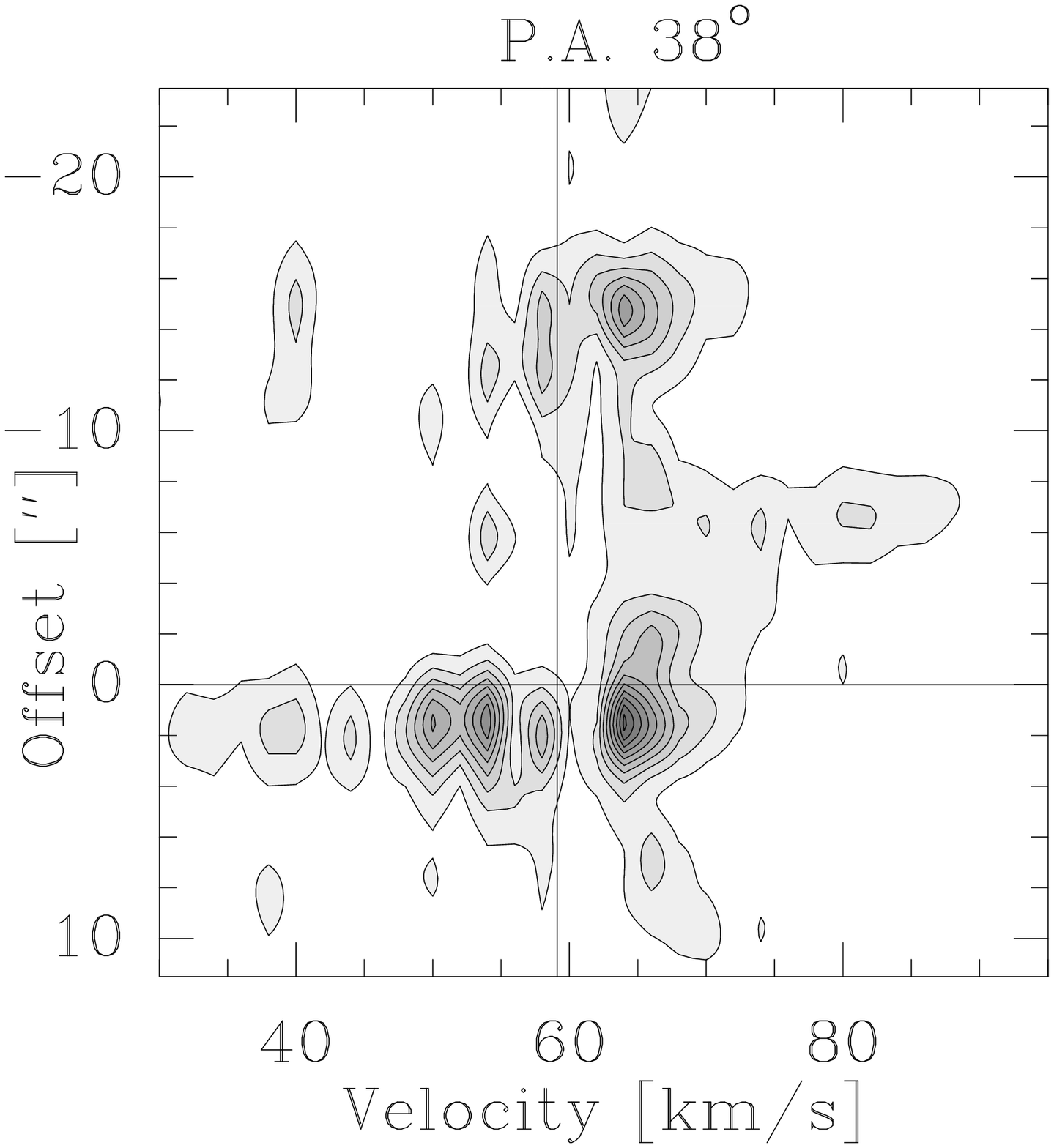} \end{center}
\caption{Position-velocity diagrams for the $^{12}$CO(2--1) SMA only
  outflow observations with a velocity resolution of 2\,km\,s$^{-1}$.
  The top panel is a pv-plot in R.A. at a Declination offset $+0.6''$
  (corresponding to the strongest features in the west). The bottom
  panel presents a pv-diagram in north-west south-eastern direction
  with a P.A of $38^{\circ}$ from north (the cuts are marked in
  Fig.\ref{sma_and_30m}, top-middle panel). The contour levels are
  from 5 to $95\%$ from the peak emission (4.8\,Jy\,beam$^{-1}$ in
  both panels).  The $v_{\rm{lsr}}$ at 59.1\,km\,s$^{-1}$ and the
  central positions are marked by vertical and horizontal lines..}
\label{pos_velo}
\end{figure}

Figure \ref{pos_velo} presents position-velocity (pv) diagrams of the
molecular gas along the directions of the two cavity walls of the
north-western outflow cone. Both diagrams show that the high-velocity
gas on the blue-shifted south-eastern side of the outflow remains very
close to the outflow center whereas the red-shifted north-western side
resembles more the typical Hubble-law with increasing velocity at
further distances from the outflow center (e.g., \citealt{arce2006}).
This very different blue/red outflow pv-characteristics are likely due
to the fact that the blue CO outflow emission apparently traces
predominantly the jet-like component also visible in cm and SiO
emission, whereas the red CO emission is found mainly toward the
cavity-like walls of the north-western outflow component.

The combined CO(2--1) data can also be used to estimate the
contributions of the two outflows emanating from IRAS\,18182$-$1433.
To calculate the outflow mass $M$ ($M_{\rm{blue}}$, $M_{\rm{red}}$,
$M_{\rm{tot}}$) the momentum $p$ and the energy $E$, we follow the
approach outlined in \citet{cabrit1990} and already employed for the
single-dish data alone by \citet{beuther2002b}. In the following
section (\S \ref{opacity}), we will discuss in more detail the varying
CO line opacities, however, for the total parameters $M/p/E$ assuming
an average $^{12}$CO/$^{13}$CO (2--1) line ratio of 10 to calculate
the opacity-corrected column densities of the blue- and red-shifted
outflow emission is a reasonable approach
\citep{choi1993,levreault1988,beuther2002b}. Furthermore, we use an
H$_2$/$^{13}$CO ratio of $89\times 10^4$ \citep{cabrit1992}, an
average temperature within the outflow of 30\,K, and the velocity
ranges of [53,57] and [62,71]\,km\,s$^{-1}$ for the integrated blue
and red outflow wings, respectively. The derived masses, momenta and
energies are shown in Table \ref{outflow_para}.

\begin{table}
\caption{Outflow parameter}
\label{outflow_para}
\begin{center}
\begin{tabular}{lrrrr}
\hline \hline
 & 30m & SMA/30m & SMA/30m(a) & SMA/30m(b)\\
\hline
$M_{\rm{blue}}$[M$_{\odot}$]   & 19.3 & 10.8 & 2.8 & 5.8 \\
$M_{\rm{red}}$[M$_{\odot}$]    & 10.8 & 7.1  & 4.5 & 2.1 \\
$M_{\rm{tot}}$[M$_{\odot}$]    & 30.1 & 17.9 & 7.3 & 7.9 \\
$p$[M$_{\odot}$\,km\,s$^{-1}$] & 246  & 151  & 71  & 60  \\
$E$[$10^{45}$erg]              & 23.0 & 14   & 8   & 5   \\
\hline
\end{tabular}
\end{center}
\end{table}

Although we recover approximately 88\% of the red and blue wing
emission when integrating over exactly the same region (see \S
\ref{shortspacings}), the derived parameters from the merged dataset
are a little bit lower because the area used to derive the outflow
parameters from the single-dish data alone is larger than for the
merged dataset (see \citealt{beuther2002b} and
Fig.~\ref{sma_and_30m}). Interestingly, both outflows contribute
approximately the same mass, momentum and energy to the total budget
of the whole quadrupolar outflow system.

\subsection{Outflow line opacities}
\label{opacity}

\begin{figure}[htb]
\begin{center}
  \includegraphics[angle=-90,width=8.8cm]{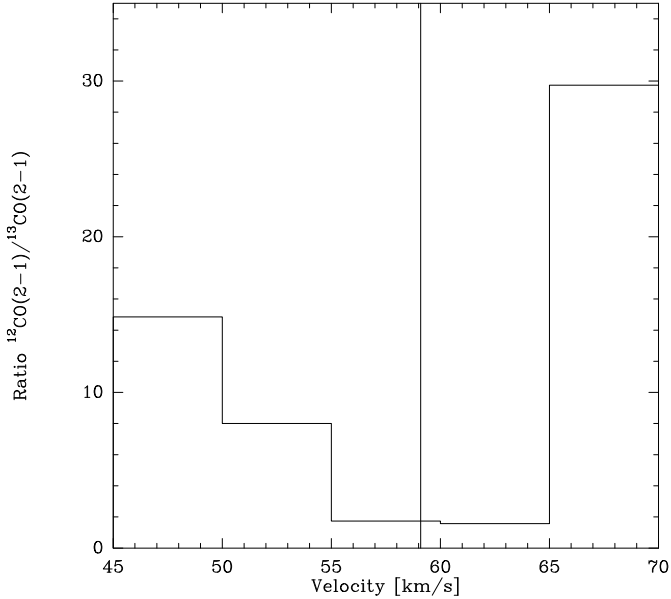}
\end{center}
\caption{Ratios between the $^{12}$CO and $^{13}$CO $J=2-1$ spectra shown 
in Fig.~\ref{spectra_co_images}. The ratio increases with increasing
velocity of the molecular outflow. We had to truncate the ratio $<45$
and $>70$\,km\,s$^{-1}$ because we detect no $^{13}$CO signal above the noise
at higher velocities. The $v_{\rm{lsr}}$ at
  59.1\,km\,s$^{-1}$ is marked by a vertical line.}
\label{ratio_spectra}
\end{figure}

The simultaneous observations of the $^{12}$CO line and their
isotopologues $^{13}$CO and C$^{18}$O (Figs.~\ref{uvspectra_co} \&
\ref{spectra_co_images}) should in principle allow to study the line
wing ratios and the corresponding CO line wing opacities with changing
velocity. The $^{12}$CO/$^{13}$CO study by \citet{choi1993} indicated
that the $^{12}$CO line opacities likely decrease with increasing
velocity with respect to the $v_{\rm{lsr}}$ but the data were not
sufficient to test this suggestion in more detail. Our new data toward
IRAS\,18182$-$1433 have the advantage that the simultaneous
observation of the three lines minimizes the calibration
uncertainties. However, the disadvantage is that the spatial filtering
effect of an interferometer can vary for the different isotopologues.
This is due to a number of reasons, two of them are: (1) Varying line
opacities between the isotopologues and thus different spatial
structures traced by each of them. (2) The synthesized maps are
affected by the side-lobes of the strongest emission features which
causes different distortion of the weaker emission in various
isotopologues. Since we have no corresponding single-dish
$^{13}$CO(2--1) and C$^{18}$O(2--1) observations, it is hard to
quantify this effect exactly for all three lines, but
\citet{beuther2005a} have shown for SMA observations of Orion-KL that
this effect can significantly change the observed line ratios.
Therefore, we stress that the absolute line ratios based on the
SMA-data alone are not reliable.  However, even the relative changes in
the $^{12}$CO(2--1)/$^{13}$CO(2--1) line ratio is interesting and
allows conclusions about the changing line-wing opacities. Figure
\ref{ratio_spectra} presents the $^{12}$CO(2--1)/$^{13}$CO(2--1) line
ratio averaged over the whole region of the molecular outflow. We
clearly see the trend of increasing line ratio with increasing
velocity. While the $^{12}$CO(2--1)/$^{13}$CO(2--1) line ratio is low
close to the $v_{\rm{lsr}}$, it increases to higher values at higher
relative velocities. Although the absolute numbers suffer from the
above discussed spatial filtering problems, it appears that the
average line wing ratio of 10 suggested by the observations of
\citet{choi1993} holds for this observations as well.

\subsection{Temperature from CH$_3$CN$(12_k-11_k)$}
\label{ch3cn}

\begin{figure}[ht]
\begin{center}
\includegraphics[angle=-90,width=8.8cm]{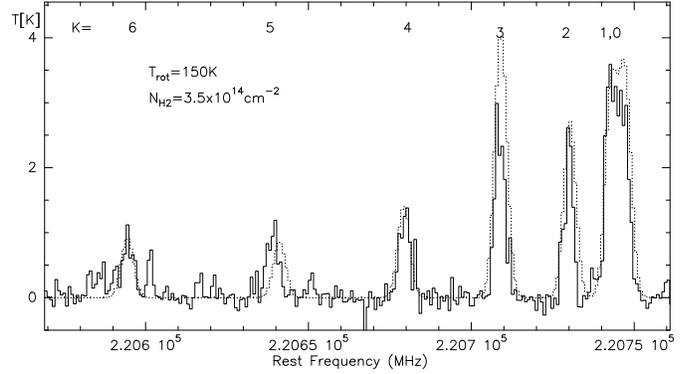}
\end{center}
\caption{The full line shows the CH$_3$CN($12_k-11_k$) spectrum toward 
the mm continuum position. The dotted line presents a model
spectrum with $T_{\rm{rot}}=150$\,K and $N_{\rm{CH_3CN}}=3.5\times
10^{14}$\,cm$^{-2}$.}
\label{ch3cn_spectrum}
\end{figure}

Since we detected 7 lines of the CH$_3$CN($12_k-11_k$) $k$-ladder with
$k=0...6$, we can utilize the varying excitation levels of the lines
with lower level energies $E_{\rm{lower}}$ between 58 and 315\,K (see
table \ref{lines}) to estimate a temperature for the central gas core.
We tried to model the whole CH$_3$CN($12_k-11_k$) spectrum toward the
mm continuum peak position in the local thermodynamic equilibrium
using the XCLASS superset to the CLASS software developed by Peter
Schilke (priv. comm.). This software package uses the line catalogs
from JPL and CDMS \citep{poynter1985,mueller2001}. Figure
\ref{ch3cn_spectrum} shows the observed CH$_3$CN($12_k-11_k$) spectrum
toward the mm peak position and a model spectrum with a temperature of
150\,K. This model spectrum fits the observation reasonable well. The
main difference is that the model spectrum is optically thin whereas
the lower line intensity of the observed $k=3$ line indicates moderate
optical depth of the CH$_3$CN lines. Nevertheless, the observations
clearly show that a hot molecular core has formed at the center of
IRAS\,18182$-$1433.

\subsection{Disk signatures?}
\label{disk}

\begin{figure*}[ht]
\begin{center}
\includegraphics[angle=-90,width=5.8cm]{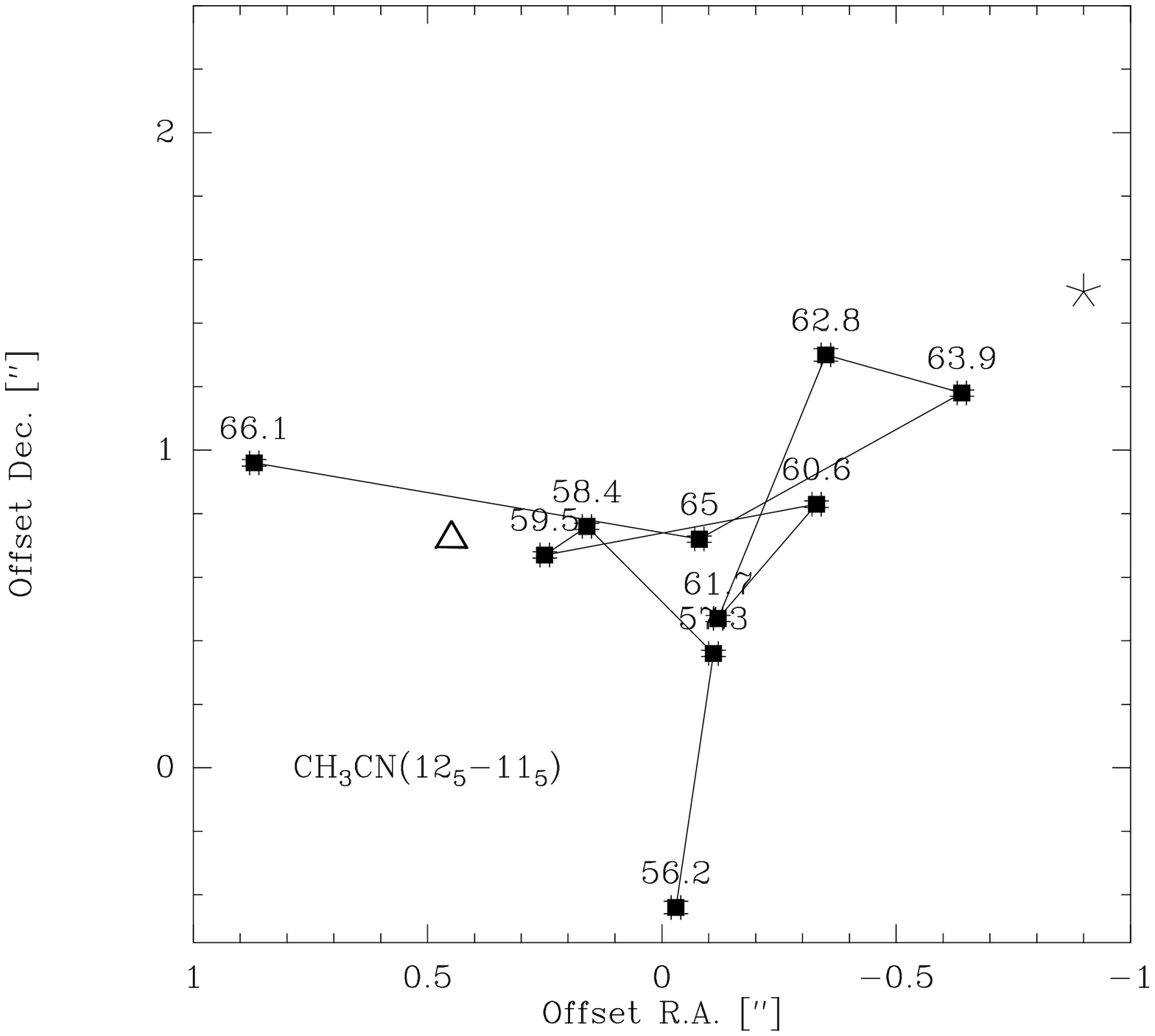}
\includegraphics[angle=-90,width=5.8cm]{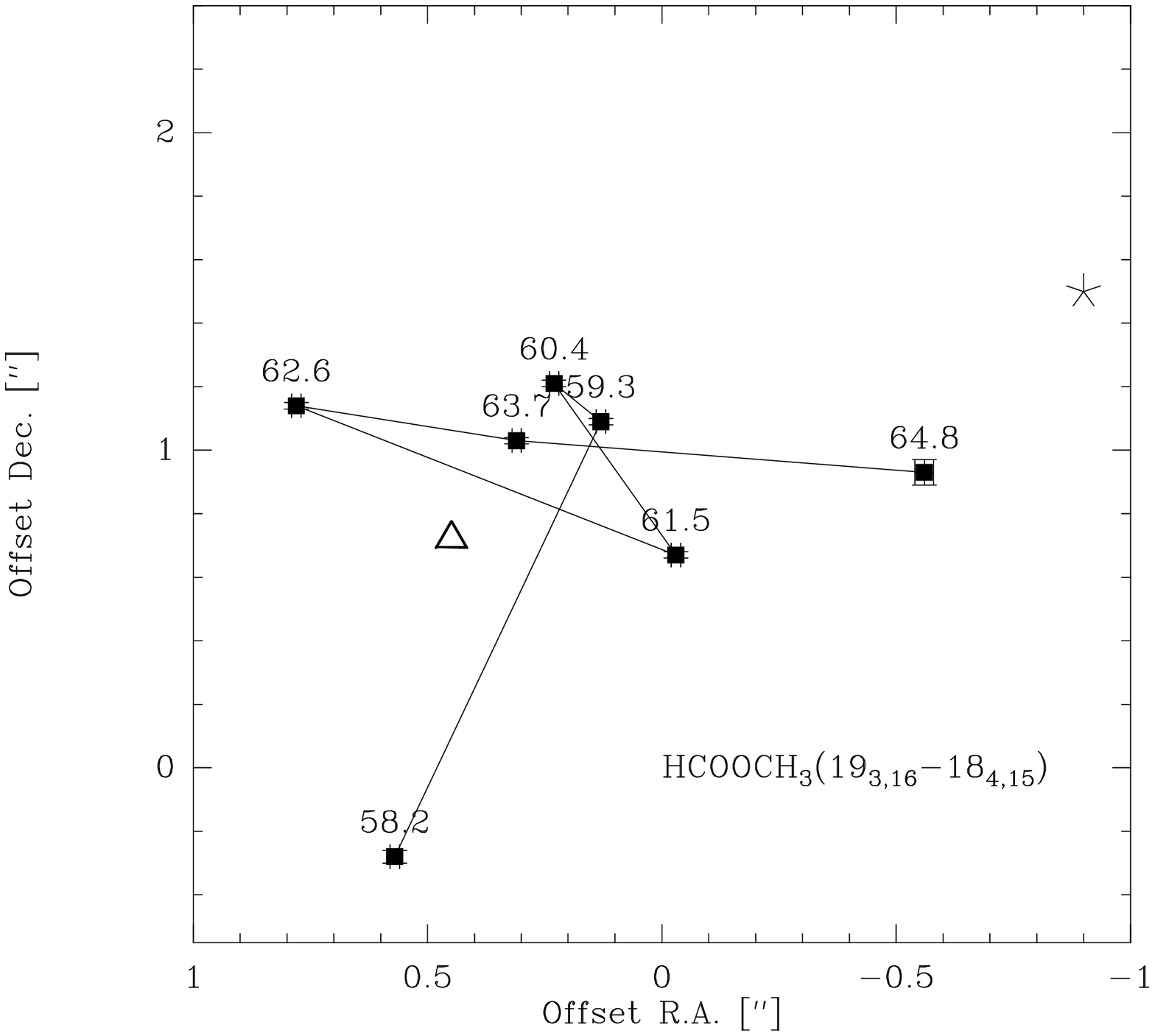}
\includegraphics[angle=-90,width=5.8cm]{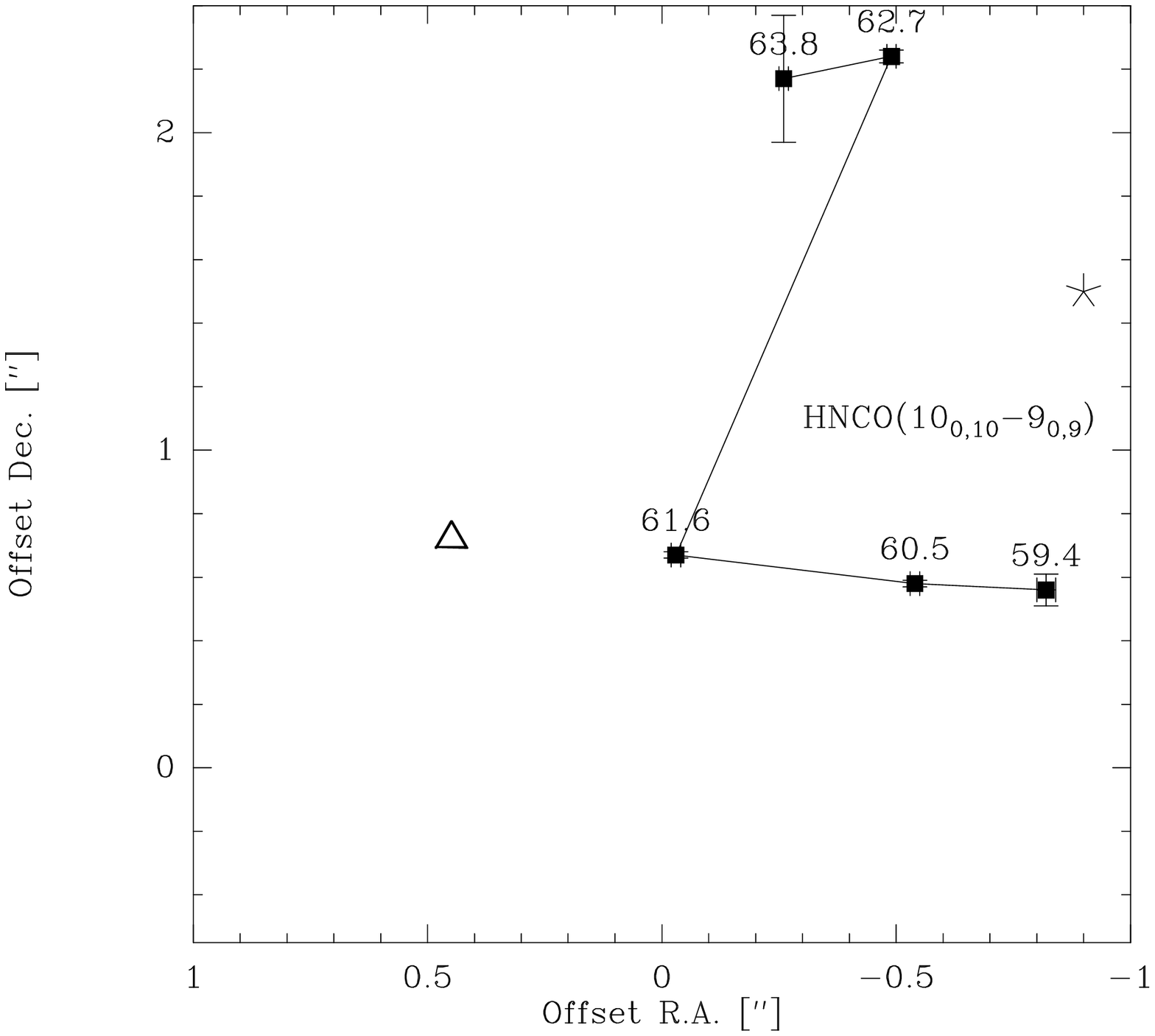}
\end{center}
\caption{Positions of different velocity channels for
CH$_3$CN$(12_5-11_5)$, HCOOCH$_3(19_{3,16})-18_{4,15})$ and
HNCO$(10_{0,10}-9_{0,9})$. The positions were fitted from the imaged
data-cubes with the task IMFIT in MIRIAD. The numbers label the center
velocities of each channel, and the error-bars are only the statistical
errors of the fitting routine. The star marks the mm continuum peak,
and the 0/0 is the phase reference center. We connected the positions
with increasing velocities to stress their random distribution.}
\label{nodisk}
\end{figure*}

Although these observations were meant to study the molecular outflow and thus
have not extremely high angular resolution ($3''$ correspond to approximately
13500\,AU at the assumed near kinematic distance), we investigated the typical
high-density hot-core tracing molecules (CH$_3$CN, HCOOCH$_3$, HNCO) for
potential disk-signatures in their velocity structures. We fitted the peak
position of each velocity channel separately to achieve a higher positional
accuracy down to 0.5\,HPBW/(S/N) \citep{reid1988} and plotted their positions
in Figure \ref{nodisk}. We cannot identify any coherent velocity
structure in these molecular lines, the velocity distributions appear to be
random and are offset from the mm continuum peak.  For CH$_3$CN and
HCOOCH$_3$, the distributions appear somehow in-between the main mm continuum
peak and the cm feature.

\section{Discussion and Conclusions}
\label{discussion}

%\subsection{Molecular Outflows}

The combined SMA+IRAM\,30\,m CO(2--1) and the VLA SiO(1--0) data
reveal a multiple outflow system emanating from IRAS\,18182$-$1433.
One of the two bipolar outflows traced by the CO(2--1) emission shows
a collimated blue and a cone-like red outflow wing. The northern
SiO(1--0) emission is suggestive of an additional third outflow
emanating toward the north. We do not identify a clear southern
bipolar counterpart of this potential third outflow.  Similar to this,
a few other examples have recently been reported where the SiO
emission is tracing only one lobe of an outflow from a massive
star-forming region (e.g., IRAS\,18089$-$1732 \citealt{beuther2004b},
IRAS\,18566+0408, \citealt{zhang2006a}). Although subject to large
uncertainties, an analysis of the $^{12}$CO(2--1)/$^{13}$CO(2--1) line
ratios indicates that the CO line wing opacities decrease with
increasing velocities with respect to the $v_{\rm{lsr}}$.

All molecular line maps show strong emission approximately $1-2''$
south-east of the mm continuum peak, well correlated with a cm
emission feature maybe due to an underlying thermal jet
\citep{zapata2006a}. The offset between the high-density tracing
molecular lines like CH$_3$CN or HCOOCH$_3$ and the 1.3\,mm continuum
peak is surprising and mainly two possibilities are likely to explain
that observational feature. First, the cm emission feature pinpoints
the location of a secondary source which remains unresolved in our mm
observations (see, e.g., IRAS\,16293$-$2422, \citealt{chandler2005}).
This secondary source then could be in an evolutionary more evolved
state altering the chemistry compared to the potentially younger
source embedded at the location of the mm continuum peak.  Contrary to
that, the cm emission could be solely due to the jet itself and not be
associated with another protostellar object (see, e.g.,
IRAS\,16547$-$4247, \citealt{rodriguez2005}). This would suggest that
the chemistry is strongly altered by the molecular outflow, which is
particularly surprising for typical hot core and dense gas tracers.
Only higher angular resolution (sub)mm continuum observations will be
capable to resolve this ambiguity.

%\subsection{Core emission}

Although our spatial resolution of $3.6''\times 2.4''$ corresponds at
the assumed near kinematic distance of 4.5\,kpc only to an average
linear resolution of $\sim 13500$\,AU, it is interesting to note that
we see no sub-clustering in the 1.3\,mm continuum data but only one
single-peaked source. This is similar to the cases of IRAS\,20126+4104
\citep{cesaroni1999} or IRAS\,18089$-$1732 \citep{beuther2005c}, but
it is different to the fragmentation of the core in IRAS\,19410+2336
\citep{beuther2004c}. However, one has to keep in mind that the
observations toward IRAS\,19410+2336 had a linear spatial resolution
of $\sim 2000$\,AU, approximately 6 times better than the new data
toward IRAS\,18182$-$1433. Therefore, it is too early to constrain
differences and similarities between the fragmentation processes of
varying cores, but as pointed out in \citet{beuther2005c}, there are
potentially different fragmentation signatures in young massive gas
cores which are worth further investigation. The identification of
three molecular outflows as well as the two cm emission features
observed in the vicinity of the mm peak implies the existence of
multiple driving sources within this central massive star-forming
core. It would be interesting to identify and characterize the driving
sources of each outflow independently, but with the given data so far
we cannot set further constrains on the individual driving sources.
Again higher angular resolution observations at (sub)mm as well as
mid-infrared wavelengths are necessary for a more accurate
characterization of the driving source properties.

%\subsection{Disk-tracing molecules}

Finding the right molecular line transitions to study massive
accretion disks remains a difficult task. The often used molecule
CH$_3$CN apparently does not work for IRAS\,18182$-$1433, but the
recently suggested HCOOCH$_3$ does not trace any rotating structure in
this source as well. While in principle it is possible that there is
no rotating gas structure in IRAS\,18182$-$1433, we consider this as
unlikely because molecular outflows require accretion and rotating
accretion disks (e.g., \citealt{cesaroni2006}).  Therefore, we believe
that the non-detection of any rotation signature close to the core
center is mainly an observational problem. Various explanations are
possible in such very young massive and dense molecular cores, the
main four are: (i) The molecular lines are usually not only produced
in accretion-disk-like rotating structures but also in their
surrounding cores, which results in various overlapping velocity
components and no coherent projected 2-dimensional velocity signature.
(ii) In many cases, the spatial resolution is still not sufficient and
multiple unresolved sources distort the observed velocity field. (iii)
Many molecular lines, even the high-density tracer, are strongly
affected by the molecular outflow distorting the observed velocity
field. (iv) We are dealing with extremely high gas column densities
and thus potentially high opacities. This way, one would not even
trace the inner rotating structure. These observations confirm that
the quest for the best disk-tracing molecular line transitions is
still open.

\begin{acknowledgements} 
  Thank you very much to Peter Schilke for providing the XCLASS software to
  model the CH$_3$CN spectra. Thanks a lot also to Yuan Chen for providing the
  integrated VLA SiO(1--0) image prior to publication. H.B.~acknowledges
  financial support by the Emmy-Noether-Program of the Deutsche
  Forschungsgemeinschaft (DFG, grant BE2578).
\end{acknowledgements}

\end{document}